%% file: EffectiveTemperatureActiveShearedFluids_2.tex
\documentclass[twocolumn,secnumarabic,amssymb, nobibnotes, aps, prl]{revtex4-1}
\usepackage{amsmath}
\usepackage{graphicx}
\usepackage{multirow}
\usepackage{array}
\usepackage[FIGTOPCAP]{subfigure}
\usepackage{graphicx}
\usepackage{tabularx}
\usepackage[usenames, dvipsnames]{color}
\usepackage{epsfig}
\usepackage{epstopdf}

\usepackage{color}

\usepackage{hyperref}

\begin{document}

\title{Effective temperature of active fluids and sheared granular matter}
\author{Saroj Kumar \surname{Nandi}}\author{N. S. \surname{Gov}}%
\email[Corresponding author: ]{nir.gov@weizmann.ac.il}
\affiliation{Department of Chemical and Biological Physics, The Weizmann Institute
of Science, Rehovot 76100, Israel}
\begin{abstract}
The dynamics within active fluids, driven by internal activity of the self-propelled particles, is a subject of intense study in non-equilibrium physics. These systems have been explored using simulations, where the motion of a passive tracer particle is followed. Similar studies have been carried out for passive granular matter that is driven by shearing its boundaries. In both types of systems the non-equilibrium motion have been quantified by defining a set of ``effective temperatures'', using both the tracer particle kinetic energy and the fluctuation-dissipation relation. We demonstrate that these effective temperatures extracted from the many-body simulations fit analytical expressions that are obtained for a single active particle inside a visco-elastic fluid. This result provides testable predictions and suggests a unified description for the dynamics inside active systems.
\end{abstract}

\pacs{}
\maketitle

\emph{Introduction.} The dynamics within dense, non-equilibrium fluids is a subject of great current interest \cite{giomi2010sheared,berthier2013non}. Such systems are driven out of equilibrium either by the constituent particles being self-propelled, such that they are moving due to internally generated forces, or by external driving such as shearing. At high densities such systems approach the glass transition, and the effects of active forces on this transition have been intensively explored. Progress in this field has relied on investigations of the granular fluids using computer simulations (active \cite{loi2008effective,loi2011effective,levis2015single}, sheared \cite{berthier2002shearing,berthier2002nonequilibrium,makse2002testing,potiguar2006effective}), and experiments (active \cite{palacci2010sedimentation}, sheared \cite{song2005experimental,PhysRevE77061309}). Analytic descriptions of the observed dynamics are scarce. The non-equilibrium dynamics in these systems is often characterized by an "effective temperature", relating the response and correlation functions, which plays an important role in the quantification of the departure from equilibrium \cite{hohenberg1989,cugliandolo1997,cugliandolo2011effective,shen2004,wang2011,wang2013,lu2006,loi2008effective}.

Recently, we have extended mode-coupling theory (MCT) and random first order transition theory (RFOT) of passive glass forming systems to that of an active system of self-propelled particles. Within MCT, we find that the system is characterized by an evolving effective temperature, which equals to the equilibrium temperature at very short time and saturates to a larger value at long time. The functional dependence of the long-time value of the effective temperature on the parameters of activity is well described by an analytic expression derived for the dynamics of a single active particle inside a caging potential, that characterizes an effective viscoelastic medium \cite{nandi2017nonequilibrium,mandal2016active}. The functional form of the increase in the effective temperature due to activity turns out to be captured by the potential energy of the particle, while the effective medium is described by an effective viscosity and elastic confining potential, which serve as fitting parameters. Within RFOT, we have obtained the correction to the configurational energy using a similar one-particle model \cite{nandi2018rfot}. This allowed us to resolve and explain the effects of activity on the dynamics and fragility of active glasses, in terms of their dependence on the characteristics of the active force correlations \cite{nandi2018rfot}.

Here we demonstrate that the same single trapped active particle approach gives an excellent description of the dependence of the kinetic energy of a passive tracer particle that is embedded in an active fluid, and in a sheared granular fluid, as obtained in previous simulation studies. This agreement allows us to explain the observed relation between the kinetic energy and the zero-frequency limit ``effective temperature" obtained from the generalized fluctuation-dissipation theorem (FDT) $T_{eff,FDT}$ \cite{cugliandolo2011effective}. Finally, we use our model to make predictions for future simulation studies.

\emph{Kinetic energy of a tracer particle in active fluids.} In the simulations of an active fluid composed of self-propelled particles (SPP) \cite{loi2008effective,loi2011effective}, the activity is often implemented by fixing the fraction of active particles, the amplitude of the force that they exert ($f_0$), and the duration of the force ("persistence time" $\tau_p$). This kind of model was used to simulate the dynamics inside an active fluid composed of spherical particles \cite{loi2008effective}, and chain-like active polymer fluid \cite{loi2011effective}.

One measure of the activity within these systems was extracted by the mean kinetic energy of a passive tracer particle that is immersed inside the active fluid \cite{loi2008effective,loi2011effective}. We therefore first write the kinetic energy that we obtain from our single-trapped active particle (STAP) model \cite{ben2015modeling}
\begin{eqnarray}
T_{\textrm{kin}}^a&=&\frac{1}{2}m_{\textrm{tr}}\langle v^2\rangle_{a}\nonumber \\
&=&\frac{p_{on}N_mf_0^2\tau_p}{4\left[\lambda\left(1+\lambda\tau_p/m_{\textrm{tr}}\right)+\tau_p^2k^{3/2}/m_{\textrm{tr}}^{1/2}\right]}
\label{ekinactive}
\end{eqnarray}
where the mass of the tracer particle is $m_{\textrm{tr}}$, which is being "kicked" along one
dimension by $N_m$ "motors" that are characterized by a fixed force $f_0$ (which can also be given by a distribution of values), mean burst length $\tau_p$ and mean waiting time between bursts
$\tau_w$, such that $p_{on}=\tau_p/(\tau_w+\tau_p)$ is the probability of a
motor to be turned on. The mean total kinetic energy of the particle is the sum of the active and thermal contributions: $T_{eff}=T+T_{\textrm{kin}}^a$.

Note that previously the expression in Eq.\ref{ekinactive} was calculated for a normalized particle mass \cite{ben2015modeling}, but we now explicitly retain the mass in the expressions, for comparison with the simulation results. When comparing the expression obtained from our STAP model to the simulations, we need to fit the parameters of the effective medium that confines the tracer particle, i.e. the effective friction coefficient ($\lambda$), and the effective elastic confinement ($k$).


We now compare the analytic expression that we obtained for the tracer particle's mean kinetic energy $T_{\textrm{kin}}^a$ (Eq.\ref{ekinactive}), with the values obtained in the simulations \cite{loi2008effective,loi2011effective}. First, we predict that $T_{\textrm{kin}}^a$ increase quadratically with the active force magnitude $f_0$, and this is indeed observed in the simulations.

\begin{figure}
\includegraphics[width=1\columnwidth]{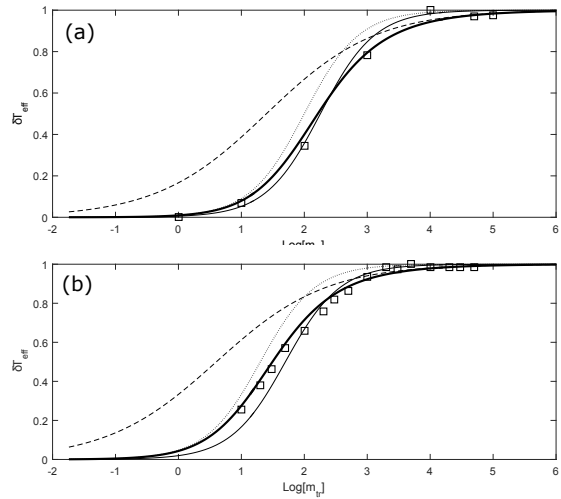}
\caption{Comparison between the normalized increase in the mean kinetic energy of the tracer particle: $\delta
T_{eff}=(T_{eff}-T)/T_{kin}(m_{\textrm{tr}}\rightarrow\infty)$, as a function
of the tracer particle's mass $m_{\textrm{tr}}$. The fits using Eq.\ref{ekinactivefit} are given by the thick and thin solid lines, using either both non-zero $A$ and $B$ ($A=20, 100$ and $B=2, 5$ for (a),(b) respectively), or only $A$ ($A=50, 180$ for (a),(b) respectively). The thin dotted and dashed lines show the contribution of the terms that depend on $A$ and $B$ alone in Eq.\ref{ekinactivefit}, respectively. The
results of numerical simulations (squares) for (a) an active fluid of spherical particles
\cite{loi2008effective}, and (b) active polymer-like chains \cite{loi2011effective}.}\label{Fig1}
\end{figure}

Next, in fig.\ref{Fig1} we plot the comparison between $T_{\textrm{kin}}^a$ and the kinetic energy of the tracer particle in te simulations \cite{loi2008effective,loi2011effective}, as function of the tracer particle's mass $m_{\textrm{tr}}$. In order to fit the data we simplify Eq.(\ref{ekinactive}) to be in the form
\begin{eqnarray}
T_{\textrm{kin}}^a=\frac{T_{\textrm{kin},0}^a}{1+A/m_{\textrm{tr}}+B/m_{\textrm{tr}}^{1/2}}
\label{ekinactivefit}
\end{eqnarray}
where $A=\lambda\tau_p$, and $B=\tau_p^2k^{3/2}/\lambda$. We find that for both of the data sets we can get reasonably good agreement by neglecting the elastic component (i.e. setting $k=B=0$) and treating the tracer as being in a purely viscous fluid, using the fit parameters $A=50, 180$ respectively. We can get a better fit, especially at the larger mass range, where the second term in the denominator of Eq.\ref{ekinactivefit} begins to dominate, using the parameters: $A=20, 100$ and $B=2, 5$ respectively. The effective elastic confinement is found to be small in these simulations, which is expected as the systems were relatively dilute, with density that is far below the jamming or glass transition values.


We further predict that the maximal value of the kinetic energy $T_{\textrm{kin},0}^a$, for $m_{\textrm{tr}}\rightarrow\infty$ (Eq.\ref{ekinactivefit}) is independent of the tracer particle's mass, as observed. The excellent agreement we obtained in Fig.1 indicates that the effective medium coefficients $\lambda,k$ are independent of $m_{\textrm{tr}}$, which is expected since in the simulations the tracer mass was varied while keeping its size constant.

\emph{Generalized Fluctuation-Dissipation Theorem in active fluids.} Another measure for the activity is obtained using the generalized Fluctuation-Dissipation Theorem (FDT), and this was extracted from the density fluctuations of the active fluid \cite{loi2008effective,loi2011effective}. From our STAP model we obtain the following expression for the generalized FDR temperature \cite{ben2015modeling}
\begin{equation}
T_{\textrm{FDT}}(\omega)=\frac{p_{on}N_mf_0^2}{4\lambda\tau_p}\frac{1}{\omega^2+\tau_p^{-2}}
\label{tefffdt}
\end{equation}
Note that this value is independent of the elastic stiffness $k$ that confines the tracer particle and of the tracer particle's mass. This expression is therefore the same as for an active particle in a purely viscous fluid \cite{ben2011effective}. In the steady-state limit (infinite time-scale, $\omega\rightarrow 0$) we get that
\begin{equation}
T_{\textrm{FDT}}(\omega\rightarrow
0)=T_{\textrm{FDT},0}=\frac{p_{on}N_mf_0^2\tau_p}{4\lambda\tau_p}
\label{tefffdt1}
\end{equation}

However, we have found in a previous study that the $T_{FDT}$ that is obtained using an extension of the mode-coupling theory (MCT) to active fluids (dense active systems of self-propelled particles) agrees with the \emph{potential energy} of the STAP model, which is given by \cite{ben2015modeling}
\begin{eqnarray}
T_{\textrm{pot}}^a&=&\frac{1}{2}k\langle x^2\rangle_{a}\nonumber \\
&=&\frac{p_{on}N_mf_0^2\tau_p}{4\lambda\left(1+k\tau_p/\lambda\right)}
\label{epotactive}
\end{eqnarray}
Note that similar to $T_{\textrm{FDT},0}$ (Eq.\ref{tefffdt1}), this expression is independent of the tracer particle's mass. However, it differs by having a dependence on $k$.

It therefore remains unclear which of these two expressions that we obtained, $T_{\textrm{FDT},0}$ or $T_{\textrm{pot}}^a$, correspond to the value of $T_{\textrm{FDT}}$ extracted from simulations. However, both analytic expressions that we obtained depend quadratically on $f_0$, and this is indeed the behavior found for $T_{\textrm{FDT}}$ in the simulations \cite{nandi2017nonequilibrium,preisler2016}.

In the simulation study, it was found that the value of the mean kinetic energy of the tracer particle agreed with the value of $T_{\textrm{FDT}}$, for the most massive tracer particles. We can calculate the ratio between the mean kinetic energy and both expressions $T_{\textrm{FDT},0},T_{\textrm{pot}}^a$, within our model:
\begin{equation}
\frac{T_{\textrm{FDT},0}}{T_{kin}}=1+\frac{\lambda\tau_p}{m_{\textrm{tr}}}+\frac{\tau_p^2k^{3/2}}{\lambda m_{\textrm{tr}}^{1/2}}
\label{tefffdtratio}
\end{equation}
and
\begin{equation}
\frac{T_{pot}}{T_{kin}}=\frac{1+\frac{\lambda\tau_p}{m_{\textrm{tr}}}+\frac{\tau_p^2k^{3/2}}{\lambda m_{\textrm{tr}}^{1/2}}}{1+\frac{k\tau_p}{\lambda}}
\label{tefffdtratio2}
\end{equation}
Note that both of these ratios approach one in the limit of $\tau_p\rightarrow0$, where there is no effect of activity and equipartition is recovered. The ratio in Eq.\ref{tefffdtratio} approaches one in the limit of  $m_{\textrm{tr}}\rightarrow\infty$, while the ratio in Eq.\ref{tefffdtratio2} approaches one in this limit when the ratio $k\tau_p/\lambda$ is vanishingly small (very weak effective confining potential compared to the viscous friction).
From the values of the fit parameters we obtained in Fig.\ref{Fig1}, for the largest tracer mass of $m_{\textrm{tr}}=10^5$, we therefore conclude that these ratios are close to $1$ in these simulations. However, we do not expect these two measures of the active motion to be the same in general, and Eqs.\ref{tefffdtratio},\ref{tefffdtratio2} predict that the deviation should increase for small tracer mass, strong confinement (large $k$) and long persistence time.

\emph{Velocity distribution and relaxation time.} The velocity distribution of the tracer particle was found in the simulations to follow a Gaussian distribution \cite{loi2008effective,loi2011effective}, for all values of
$m_{\textrm{tr}}$. From our analysis of the model \cite{ben2011effective,ben2015modeling} we expect the velocity distribution to be Gaussian in either one of the two following cases: (i) when $\lambda\tau_p/m_{\textrm{tr}}\ll1$, which is indeed satisfied in the simulations for the largest tracer masses (using the parameters we fitted in Fig.\ref{Fig1}), and (ii) when $\lambda\tau_p/m_{\textrm{tr}}\gg1$ but the tracer particle is being "kicked" simultaneously by a large number of active motors, i.e. $N_m\gg1$. Within the simulations the tracer particle can be in contact with several neighboring active particles that affect it, which may give rise to the observed Gaussian distribution.

\begin{figure}[b]
\subfigure[]{
\includegraphics[scale=.5]{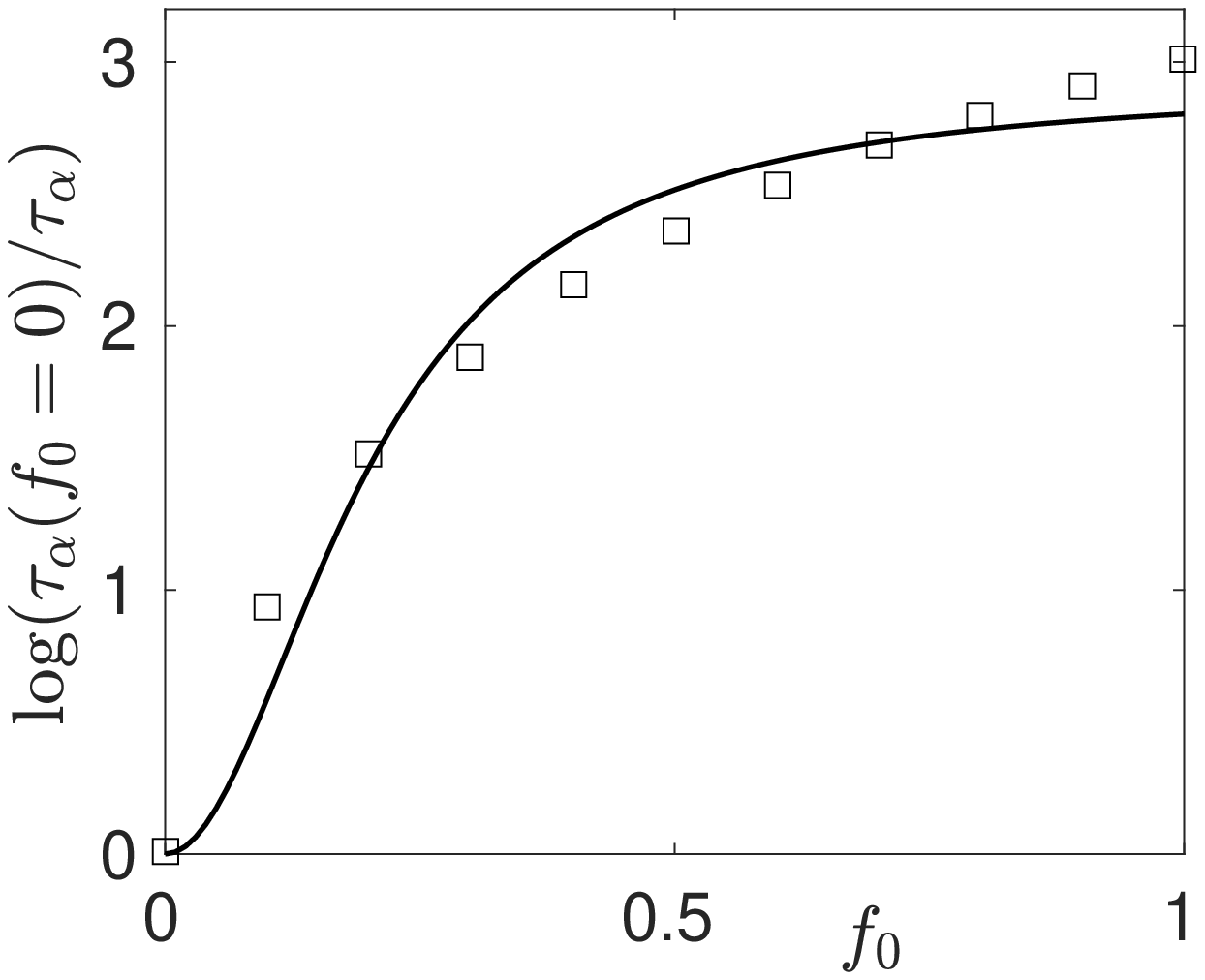}
\label{fig:figure2a}
}
\newline
\subfigure[]{
\includegraphics[scale=.5]{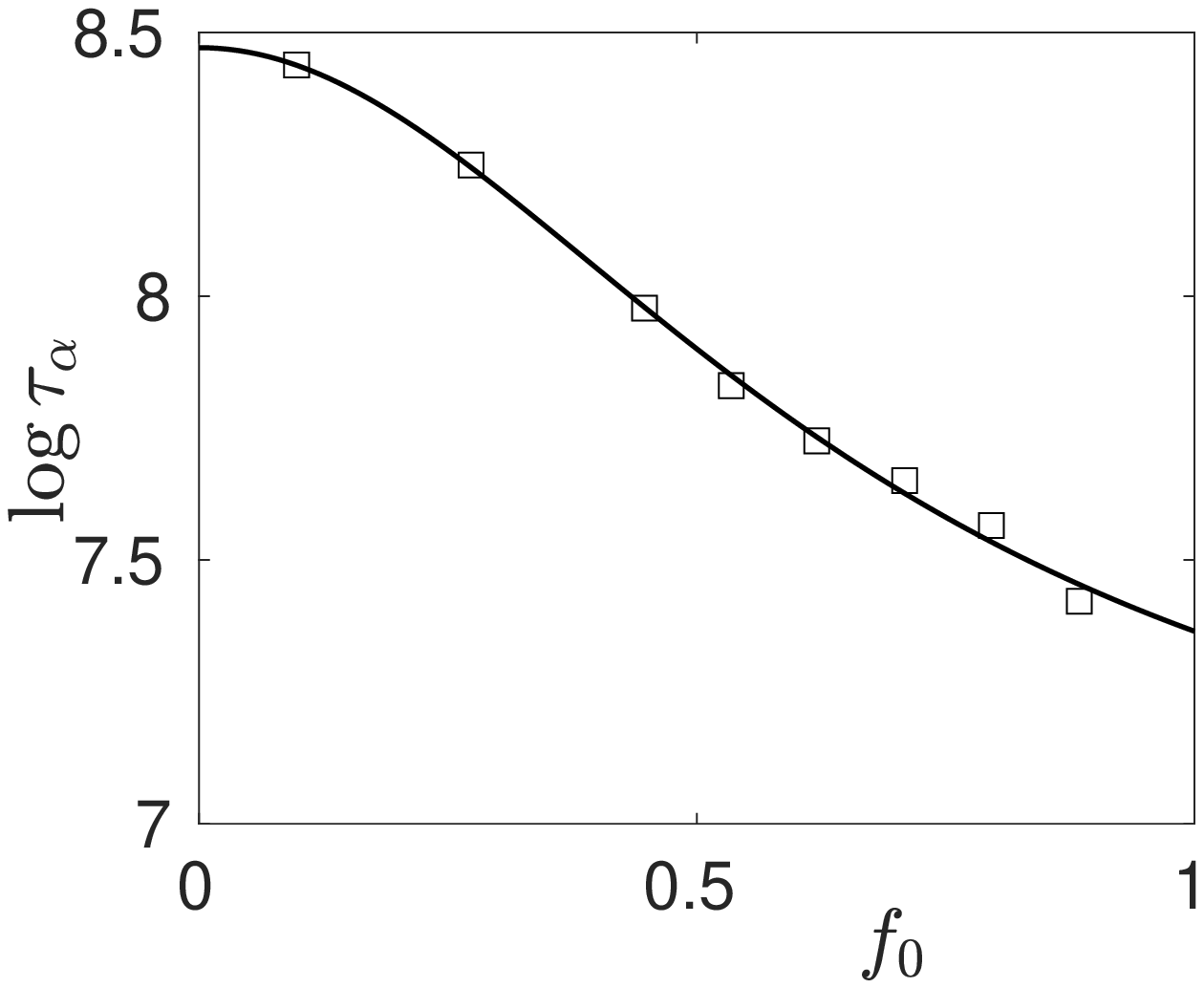}
\label{fig:figure2b}
}
\caption{Comparison between the analytic expressions (dashed lines) for $\tau_{\alpha}$ (Eqs.\ref{fit_softmat}),\ref{fit_pre}), and the simulations (filled circles) of an active fluid from \cite{loi2011effective,loi2008effective}, (a) and (b) respectively). The fitting parameters are given in the text.}
\label{fig:figure2}
\end{figure}

Finally, the relaxation dynamics was quantified in the simulations through the temporal decay of the incoherent (one-particle) intermediate scattering function. This defines the $\alpha$-relaxation time $\tau_{\alpha}$, which was found to decrease for increasing motor force ($f_0$) \cite{loi2008effective,loi2011effective}. We can relate the relaxation time with the effective temperature through a simple Arhenius-like process through
\begin{equation}
\log{[\tau_{\alpha}/\tau_0]}=\frac{E}{T_{eff}}\label{arhenius}
\end{equation}
where $E$ is an energy scale. Then for the passive system, we should have $\log[\tau_\alpha(f_0=0)]=E/T$. Therefore, using Eq.\ref{arhenius} above, we obtain
\begin{equation}\label{fit_softmat}
 \log\left[\frac{\tau_\alpha(f_0=0)}{\tau_\alpha}\right]=\frac{\tilde{\Gamma}f_0^2}{1+\Gamma f_0^2}
\end{equation}
where we have written $T_{eff}=1+\Gamma f_0^2$ and $\tilde{\Gamma}=\Gamma E$ and used $T=1$ as in the simulation. Using $\tilde{\Gamma}$ and $\Gamma$ as
fitting parameters, and fitting Eq. (\ref{fit_softmat}) with the data obtained from Fig. 12 of \cite{loi2011effective}, we obtain
$\tilde{\Gamma}=73.63$ and $\Gamma=25.27$ and plot the data along with Eq. (\ref{fit_softmat}) in Fig. \ref{fig:figure2}a. Similarly, for the data in \cite{loi2008effective}, we use the form
\begin{equation}\label{fit_pre}
 \log\tau_\alpha=\Gamma_1+\Gamma_2/(1+\Gamma_3 f_0^2)
\end{equation}
where $\Gamma_1=6.86$, $\Gamma_2=1.61$ and $\Gamma_3=2.2$ and show the plot of Eq. (\ref{fit_pre}) and the data from Fig. 2 of \cite{loi2008effective} in Fig. \ref{fig:figure2}b.

\emph{Comparison with sheared granular simulations.} Unlike the active fluids, where the self-propelled particles are driven by active noise, in a sheared granular fluid the energy is supplied to the particles from the shearing motion of the boundaries. In this system, the forces that kick the particles arise from the shearing motion of the boundaries, which cause stresses to buildup and distribute throughout the granular material. The resulting network of force chains drives local rearrangements of the grains that release the built-up stresses, and drive the motion of the particles \cite{majmudar2005contact}.

The mean kinetic energy of the particles, as well as the $T_{FDR}$, of a sheared granular fluid were extracted from simulations \cite{berthier2002nonequilibrium,berthier2002shearing}, and experiments \cite{losert2000particle,song2005experimental}. The detailed
mechanism of how shear and activity drive the systems
out of equilibrium are different. However, since both $\tau_p$ and the shear rate dictate the temporal correlations of the external drive, they should be related in some way. It is not completely clear how to relate the activity parameters $f_0,\tau_p$ of the STAP model to the shear-rate used in the simulation. The events that convert the internal built-up stress to motion involve local rearrangements of the particles, and their average duration is related to the parameter $\tau_p$ of the kicked-particle model. We will assume that the effective $\tau_p$ is largely independent of $m_{tr}$.
This means that the dependence of the kinetic $T_{eff}$ on the mass of the tracer particle should be captured by Eq.\ref{ekinactivefit}, even for a sheared system.

\begin{figure}
\includegraphics[width=1\columnwidth]{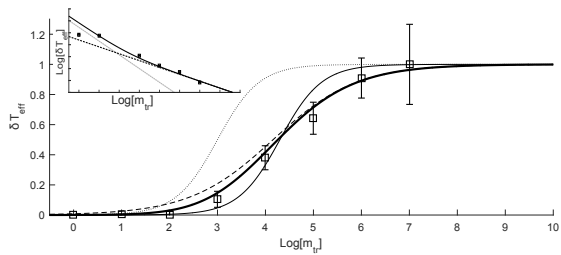}
\caption{Comparison between the normalized increase in the mean kinetic energy of the tracer particle: $\delta
T_{eff}=(T_{eff}-T)/T_{kin}(m\rightarrow\infty)$, as a function
of the tracer particle's mass $m_{\textrm{tr}}$. The fits using Eq.\ref{ekinactivefit} are given by the thick and thin solid lines, using either both non-zero $A$ and $B$ ($A=2*10^3$ and $B=120$), or only $A$ ($A=2*10^4$). The thin dotted and dashed lines show the contribution of the terms that depend on $A$ and $B$ alone in Eq.\ref{ekinactivefit}, respectively. The
results of numerical simulations (squares) for sheared granular system \cite{berthier2002shearing}. The inset shows a log-log plot, to emphasize the clear dominance of the $m_{\textrm{tr}}^{-1/2}$ term for large masses.}\label{Fig3}
\end{figure}

In fig.\ref{Fig3} we plot the comparison between Eq.\ref{ekinactivefit}) and the mass dependence of the mean kinetic energy of the tracer particle in a sheared granular system \cite{berthier2002shearing}. We find excellent agreement, this time with much larger component of elastic confinement, as compared to the dilute active fluids shown in Fig.\ref{Fig1}.

In \cite{berthier2002shearing,berthier2002nonequilibrium} it was found that there is an
identity between the kinetic energy and $T_{FDR}$ for
$m_{\textrm{tr}}\geq10^6$. From our fit in Fig.\ref{Fig3} we find that
in these simulations $\lambda\tau_p/m_{tr}\ll1$ for the largest mass of the tracer
particle, and from Eqs.\ref{tefffdtratio},\ref{tefffdtratio2} we expect that the kinetic and FDR temperature will be almost equal in this regime. As for the active fluids, within our model we do not in general expect these different measures of activity to be the identical \cite{potiguar2006effective}.

\emph{Conclusion.} We have shown that the analytical expression for the kinetic energy that we obtain from the STAP model gives an excellent description of the kinetic energy of a tracer particle immersed in an active fluid or in a sheared granular systems, as obtained previously using numerical simulations. This result highlights that the dynamics within the many-body active system can be captured using the calculation of a single particle moving within an effective medium. Note that our assumption that the effective medium properties ($\lambda,k$) of the effective single particle scenario are independent of the activity could breakdown at large activity. Specifically, our theory may not be applicable when there is activity-induced phase separation.

The excellent agreement that we obtained between the many-body simulations and the calculation for a single trapped particle may seem surprising at first: the tracer particle in the simulations performs diffusion over long times ($\gg\tau_{\alpha}$), so why should a trapped particle picture be applicable ? The agreement arises from the fact that within the many-body system the tracer particle spends most of its time within local potential minima, while the transitions between these minima occur over a relatively short time ($\ll\tau_{\alpha}$). Therefore, when calculating the mean kinetic and potential energies of the particle, the time spent within the local minima dominate.

Our model provides many testable predictions, such as:
(i) We can predict the dependencies of the kinetic energy and FDR temperature on the size (radius) of the tracer particle $R$. The following quantities that enter Eq.(\ref{ekinactive}) depend on $R$: The mass of the tracer particle  $m_{\textrm{tr}}\propto R^d$ (where $d$ is the dimensionality of the system), the number of simultaneous active particles that the tracer particle interacts with grow as: $N_m\propto R^{d-1}$, and the friction coefficient: $\lambda\propto R^\mu$, where $1<\mu\leq d-1$. In 3D, for viscous friction ($\mu=1$), we therefore get from Eqs.\ref{ekinactive},\ref{ekinactivefit} that: $T_{\textrm{kin}}^a\propto R/(1+A/R^3+B/R^{3/2})$, which increases with increasing tracer size.


(ii) We predict that for more persistent active articles, with larger $\tau_p$, there will be a growing discrepancy between the two measures of activity, such that the ratio $T_{\textrm{FDT}}/T_{kin}$ increases.

(iii) We also predict that the mean kinetic energy (Eq.\ref{ekinactive}) will be an increasing function of the persistence $\tau_p$, for small $\tau_p$, but decreasing for large $\tau_p$. This prediction applies for a model of the active noise with constant active force $f_0$ \cite{nandi2017nonequilibrium}. However, if the self-propelled particle activity has temporal correlations that approach a $\delta$-function as $\tau_p\rightarrow0$ \cite{flenner2016nonequilibrium}, such that $f_0^2\sim T_{m}/\tau_p$ (where $T_m$ is a constant) \cite{nandi2018rfot}, we predict that the mean kinetic energy decreases with increasing persistence time.

These predictions await future numerical and experimental studies. Such tests could define the limitations of the proposed analogy, as well as expose the relation between the effective single-particle parameters and the actual microscopic properties (such as density, particle interactions etc.) of the many-body system.

\begin{acknowledgments}
{\bf Acknowledgments:} N.S.G. is the incumbent of the Lee and William Abramowitz Professorial Chair of Biophysics. This research is made possible in part by the generosity of the Harold Perlman Family.

\end{acknowledgments}

\input{EffectiveTemperatureActiveShearedFluids_2.bbl}
\end{document}

%% file: EffectiveTemperatureActiveShearedFluids_2.bbl
%